\documentclass[draftcls, onecolumn]{IEEEtran}
\usepackage{cite}      % Written by Donald Arseneau
\usepackage{graphicx}  % Written by David Carlisle and Sebastian Rahtz

\usepackage{subfigure} % Written by Steven Douglas Cochran

\usepackage{amsmath}   % From the American Mathematical Society
\usepackage{array}
\begin{document}
%
% paper title
\title{Near-Field Focusing Plates and Their Design}
%
%
% author names and IEEE memberships
% note positions of commas and nonbreaking spaces ( ~ ) LaTeX will not break
% a structure at a ~ so this keeps an author's name from being broken across
% two lines.
% use \thanks{} to gain access to the first footnote area
% a separate \thanks must be used for each paragraph as LaTeX2e's \thanks
% was not built to handle multiple paragraphs

\author{Anthony~Grbic,~\IEEEmembership{Member,~IEEE,}
        and~Roberto~Merlin,~\IEEEmembership{Fellow,~OSA}
        % <-this % stops a space
\thanks{This work was supported by the Air Force Office of Scientific Research (AFOSR) under grants FA9550-07-1-0029 and FA9550-06-01-0279 through the MURI Program.}% <-this % stops a space
\thanks{A. Grbic is with the Radiation Laboratory, Department of Electrical Engineering and Computer Science, University of Michigan, Ann Arbor, 48109-2122, USA (e-mail: agrbic@umich.edu).}
\thanks{R. Merlin is with the FOCUS Center and Department of Physics, University of Michigan, Ann Arbor,
48109-1040, USA (e-mail: merlin@umich.edu).}}

% note the % following the last \IEEEmembership and also the first \thanks -
% these prevent an unwanted space from occurring between the last author name
% and the end of the author line. i.e., if you had this:
%
% \author{....lastname \thanks{...} \thanks{...} }
%                     ^------------^------------^----Do not want these spaces!
%
% a space would be appended to the last name and could cause every name on that
% line to be shifted left slightly. This is one of those "LaTeX things". For
% instance, "A\textbf{} \textbf{}B" will typeset as "A B" not "AB". If you want
% "AB" then you have to do: "A\textbf{}\textbf{}B"
% \thanks is no different in this regard, so shield the last } of each \thanks
% that ends a line with a % and do not let a space in before the next \thanks.
% Spaces after \IEEEmembership other than the last one are OK (and needed) as
% you are supposed to have spaces between the names. For what it is worth,
% this is a minor point as most people would not even notice if the said evil
% space somehow managed to creep in.
%
% The paper headers
\markboth{submitted to IEEE Transactions on Antennas and
Propagation}{Shell \MakeLowercase{\textit{et al.}}: Bare Demo of
IEEEtran.cls for Journals}
%\MakeLowercase{\textit{et al.}}: Bare Demo of IEEEtran.cls for
%Journals}
% The only time the second header will appear is for the odd numbered pages
% after the title page when using the twoside option.
%
% *** Note that you probably will NOT want to include the author's name in ***
% *** the headers of peer review papers.                                   ***

% If you want to put a publisher's ID mark on the page
% (can leave text blank if you just want to see how the
% text height on the first page will be reduced by IEEE)
%\pubid{0000--0000/00\$00.00~\copyright~2002 IEEE}

% use only for invited papers
%\specialpapernotice{(Invited Paper)}

% make the title area
\maketitle

\begin{abstract}
This paper describes the design of near-field focusing plates, which
are grating-like structures that can focus electromagnetic radiation
to spots or lines of arbitrarily small subwavelength dimension. A
general procedure is outlined for designing a near-field plate given
a desired image, and its implementation at microwave frequencies is
discussed. Full-wave (method of moments) simulations clearly
demonstrate the near-field plate's ability to overcome the
diffraction limit. Finally, it is shown that performance of
near-field plates is weakly affected by losses.
\end{abstract}

\begin{keywords}
near-field, evanescent waves, focusing, lens, metamaterials.
\end{keywords}
% Note that keywords are not normally used for peerreview papers.

% For peer review papers, you can put extra information on the cover
% page as needed:
% \begin{center} \bfseries EDICS Category: 3-BBND \end{center}
%
% For peerreview papers, inserts a page break and creates the second title.
% Will be ignored for other modes.
\IEEEpeerreviewmaketitle

\section{Introduction}

\PARstart{T}{he} electromagnetic near-field has intrigued scientists
and engineers alike since the time Synge first proposed detecting
the near-field to obtain resolutions beyond the Abb\'e diffraction
limit \cite{Synge}. Synge showed that probing the near-field
(evanescent waves) of an object amounted to tapping into the
object's subwavelength details. Nearly fifty years after Synge's
proposal, Ash and Nicholls experimentally verified that near-field
imaging was possible in 1972 \cite{AshNicholls}. By scanning the
near-field of an object they were able to achieve a $\lambda / 60$
resolution at microwave frequencies. Experimental verification at
visible wavelengths was subsequently demonstrated in the 1980's
\cite{Pohl,Lewis}.

In addition to detecting the near-field, manipulating and focusing
it have attracted significant attention in recent years. Much of
this current interest stems from the perfect lens concept introduced
by Pendry in 2000 \cite{Pendry}. Pendry showed that a planar slab of
negative refractive index material can manipulate the near-field in
such a way that it achieves perfect imaging i.e., a perfect
reconstruction of the source's near and far-field. He also showed
that the near-field could be focused with only a negative
permittivity slab. The experimental verification of negative
refraction \cite{Shelby} and subwavelength focusing using negative
refractive index \cite{grbicPRL} and negative permittivity slabs
\cite{Zhang, Blaikie}, have demonstrated that near-field lenses are
in fact a reality.

In \cite{Roberto}, an alternative method to manipulate and focus the
near-field was proposed, which relies on a patterned (grating-like)
planar structure. The proposed planar structure will be referred to
as a near-field focusing plate. The near-field plates described in
\cite{Roberto} are capable of focusing electromagnetic radiation to
spots or lines, of arbitrarily small subwavelength size. These
plates can be thought of as impedance sheets that have a modulated
surface reactance. The sheets focus the field of a plane wave
incident from one side to the other side with subwavelength
resolution.

This paper describes the design of near-field focusing plates. It
explains how the field at the exit face of such a near-field plate
can be derived given a desired image at the near-field plate's focal
plane. In addition, a procedure is outlined for designing a
near-field plate to achieve a specific image. Finally, a microwave
implementation is discussed. Full-wave (method of moments) results
are presented that clearly demonstrate the plate's ability to
overcome the diffraction limit, that is, to focus microwaves to
subwavelength dimensions.

\section{Finding the Fields at the Near-Field Focusing Plate}

Throughout this paper it will be assumed that the near-field plate
is located along the $z=0$ plane and extends in the $x$ and $y$
directions as shown in Figure \ref{Figure1}. Furthermore, it will be
assumed that the focal plane of the near-field plate is located at
$z=L$. In order to design a near-field plate, we must first select
what image we desire to have at the focal plane. From the image, we
can then proceed to derive the fields that must be present at the
surface of the near-field focusing plate. First, a Fourier transform
is taken of the image $f(x, y,z=L)$ to obtain its plane-wave
spectrum $F(k_x, k_y, z=L)$:
\begin{figure}
\begin{center}
\includegraphics[width= 4cm]{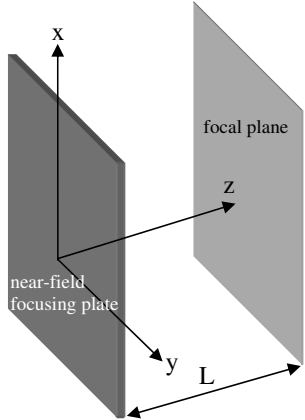}
\end{center}
\caption{Near-field focusing plate $(z=0)$ and focal plane $(z=L)$.
A near-field focusing plate is a patterned, planar structure that
can focus electromagnetic radiation to spots or lines of arbitrary
subwavelength dimension.} \label{Figure1}
\end{figure}
\begin{equation}
F(k_x, k_y,z=L) =  \int^{\infty}_{-\infty} \int^{\infty}_{-\infty}
f(x', y', L)e^{+j(k_{x'}x'+k_{y'}y')}\,dx' dy' \label{eq: eq1}
\end{equation}
where the harmonic time dependence is $e^{j\omega t}$. The
plane-wave spectrum of the image is then back-propagated to the
plane of the near-field plate, located at $z=0$:
\begin{equation}
F(k_x, k_y,z=0) =  F(k_x, k_y,L)e^{+j(k_{x}x+k_{y}y+k_{z}L)}
\label{eq: eq2}
\end{equation}
where
\begin{equation}
\begin{array}{l}
k_{z} = \sqrt{{k_\circ}^2 - {k_x}^2 - {k_y}^2} \; \; \mathrm{when} \; \; {k_x}^2 + {k_y}^2 < k_\circ^2 \\
k_{z} = -j \sqrt{{k_x}^2 + {k_y}^2 - {k_\circ}^2} \; \; \mathrm{when} \; \; {k_x}^2 + {k_y}^2 > k_\circ^2 \\
\label{eq: eq2b}
\end{array}
\end{equation}
and $k_\circ$ is the wavenumber in free space.  Back-propagation
refers to the process of reversing the phase of the propagating
plane-wave spectrum and growing (restoring) the evanescent
plane-wave spectrum, in order to recover the complete plane-wave
spectrum at the near-field plate ($z=0$). Finally, summing the
plane-wave spectrum at $z=0$ one recovers the field at the
near-field focusing plate:
\begin{equation}
f(x, y,z=0) =  \frac{1}{4 \pi^2} \int^{\infty}_{-\infty}
\int^{\infty}_{-\infty} F(k_x, k_y, L)
e^{+j(k_{x}x+k_{y}y+k_{z}L)}\, dk_x dk_y \label{eq: eq3}
\end{equation}
Given the field at the surface of the near-field focusing plate, one
can then proceed to design the plate itself. The design process
involves finding a surface impedance that yields the desired field
at the near-field focusing plate.

\section{A Near-field Focusing Plate in Two Dimensions}

\begin{figure}
\begin{center}
\includegraphics[height= 3cm]{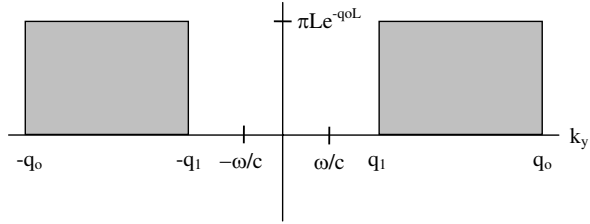}
\end{center}
\caption{The plane-wave spectrum of the image given by Equation
(\ref{eq: eq4}).} \label{Figure2}
\end{figure}

\begin{figure}
\centering \subfigure[The electric field at the near-field focusing
plate: $f(y,z=0)$.]{
    \includegraphics[width=7cm]{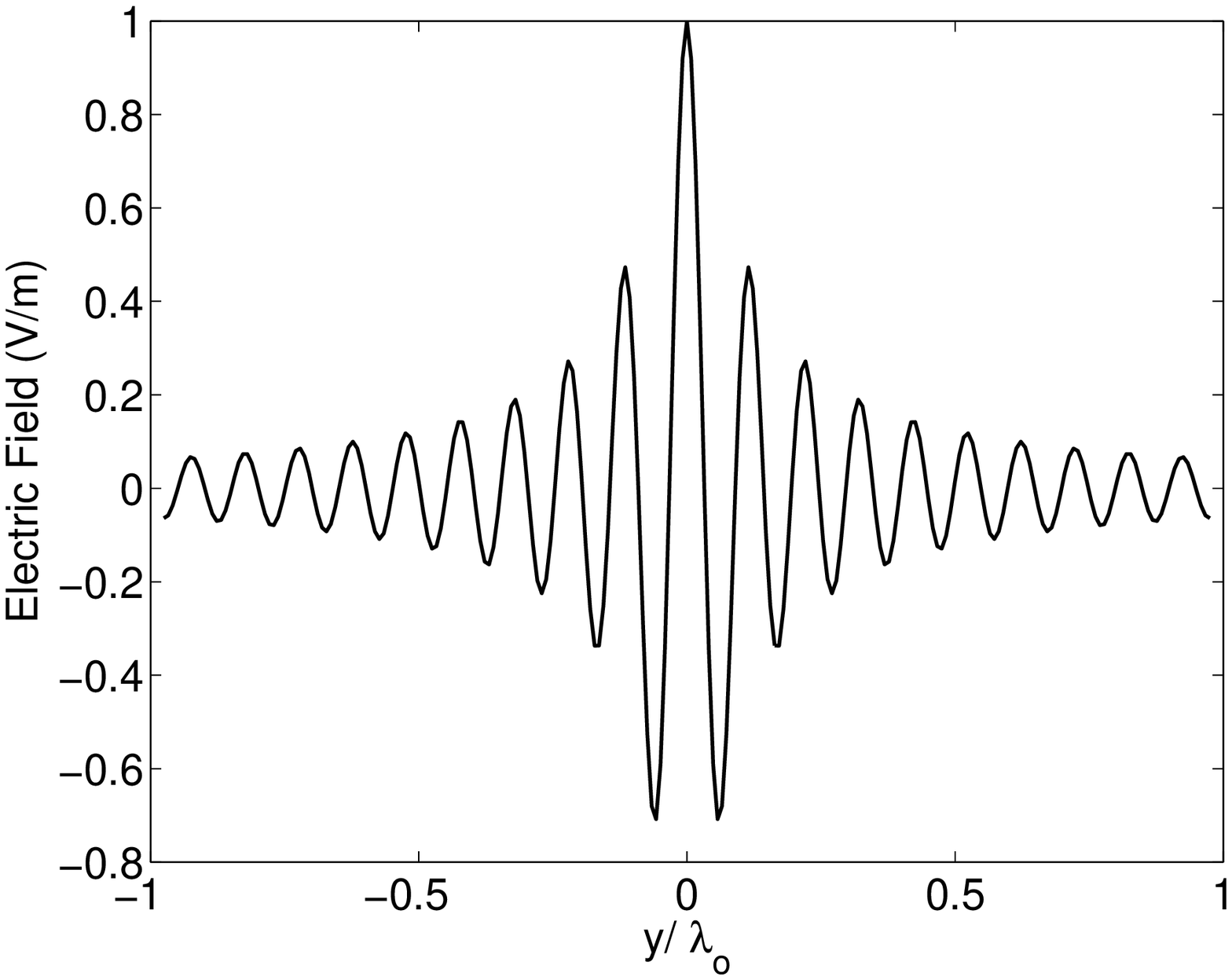}
    \label{Etotal}}
\hfill \subfigure[The electric field at the focal plane:
$f(y,z=L)$.]{
    \includegraphics[width=7cm]{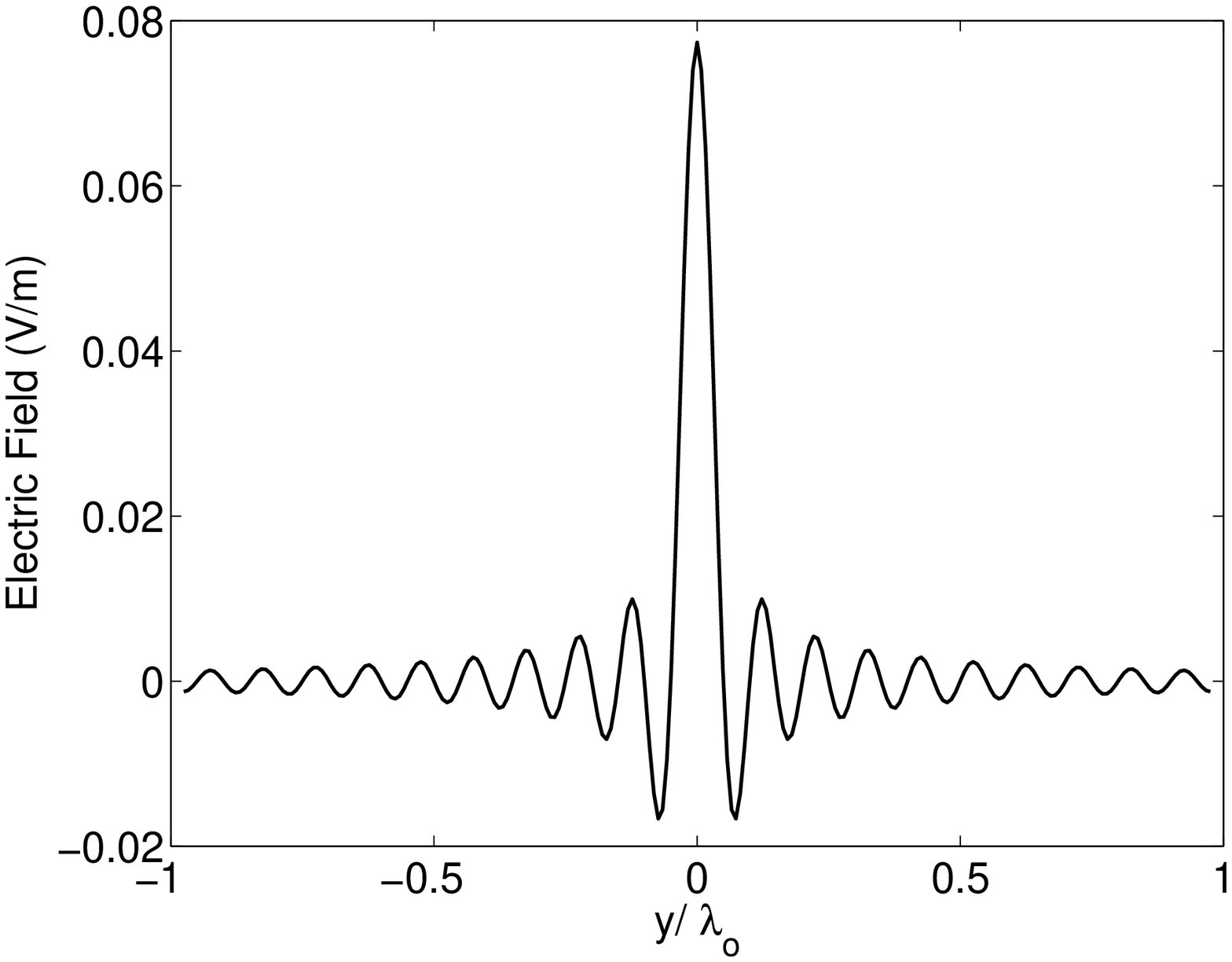}
    \label{Image}}
    \caption{Electrical field profiles at the surface of the near-field focusing plate $(z=0)$ and at the focal plane (z=L).}
    \label{Figure3}
\end{figure}

As discussed in \cite{Roberto}, various near-field plates can be
envisioned that produce focal patterns of various types and
symmetries. Although the design procedure outlined in this paper is
quite general, we will consider a simple near-field plate that
focuses evanescent waves in two dimensions ($y$ and $z$). The $y$
coordinate will denote the direction transverse to the near-field
focusing plate and $z$ the direction normal to the surface of the
plate (see Figure \ref{Figure1}). For this particular design, the
image along the focal plane ($z=L$) is chosen to be a sinc function
of the following form:
\begin{equation}
f(y,z=L) = e^{-q_0 L}q_0 L\mbox{sinc}(q_0 y) - e^{-q_0 L}q_1L
 \mbox{sinc}(q_1y) \label{eq: eq4}
\end{equation}
where $\mbox{sinc}(\theta)=\mbox{sin}(\theta)/ \theta$ and $|q_0|
\gg|q_1| \gg |k_\circ|$, $k_\circ$ is the wavenumber in free space.
The image given by Equation (\ref{eq: eq4}) has a flat
evanescent-wave spectrum of magnitude $\pi L e^{-q_0 L}$ that
extends between $q_1<|k_y|<q_0$, as depicted in Figure
\ref{Figure2}.
Such an image could be expected when imaging a line source with a
negative permittivity slab (a silver superlens) \cite{Pendry, Zhang,
Blaikie}. The propagating spectrum is zero at the focal plane since
it is totally reflected by the negative permittivity slab, but the
evanescent spectrum is still present. The spatial frequency
$k_y=q_0$ represents a cut-off wavenumber, above which transmission
through the slab rapidly falls off. However, instead of having the
evanescent spectrum fall off as in focusing using a negative
permittivity slab, we have simply assumed that it is truncated
beyond $k_y=q_0$. This cut-off wavenumber $k_y=q_0$ is dictated by
the inherent losses of the negative permittivity slab
\cite{SmithRes, Merlina}. Under the condition that $q_0 \gg q_1$ the
image simplifies to:
\begin{equation}
f(y,z=L) \approx e^{-q_0 L}q_0 L \mbox{sinc}(q_0 y) \label{eq: eq5}
\end{equation}
To find what field distribution $f(y,z=0)$ is needed at the
near-field plate to produce such an image, we back-propagate the
plane-wave spectrum of the image and then sum it up at $z=0$:
\begin{equation}
f(y,z=0) =  \frac{1}{2 \pi} \int^{-q_1}_{-q_0} \pi L e^{-q_0 L}
e^{+jk_zL}e^{+jk_xx}\,dk_x +  \frac{1}{2 \pi} \int^{q_0}_{q_1} \pi L
e^{-q_0 L} e^{+jk_zL}e^{+jk_yy}\,dk_y \label{eq: eq6}
\end{equation}
Since we are in the subwavelength region ($|q_0| \gg |q_1| \gg
|k_\circ|$) $k_z \approx -j|k_y|$. Therefore, Equation (\ref{eq:
eq6}) can be expressed as,
\begin{equation}
f(y,z=0) \approx  \frac{1}{2 \pi} \int^{-q_1}_{-q_0} \pi L e^{-q_0
L} e^{|k_y|L}e^{+jk_yy}\,dk_y + \frac{1}{2 \pi} \int^{q_0}_{q_1} \pi
L e^{-q_0 L} e^{|k_yL|}e^{+jk_yy}\,dk_y \label{eq: eq7}
\end{equation}
Performing the above integration, the following expression is
obtained for the field at the surface of the near-field plate:
\begin{equation}
f(y,z=0) \approx \frac{L[L \cos(q_0 y)+y \sin(q_0 y)] -
e^{(q_1-q_0)L}[L \cos(q_1y)+y \sin(q_1y)]}{[L^2+y^2]} \label{eq:
eq8}
\end{equation}
Given that $q_0 \gg q_1$, this expression simplifies to:
\begin{equation}
f(y,z=0) \approx \frac{L[L \cos(q_0 y)+y \sin(q_0 y)]}{[L^2+y^2]}
\label{eq: eq9}
\end{equation}
From Equations (\ref{eq: eq5}) and (\ref{eq: eq9}), it is apparent
that the field at the near-field plate decays toward the focal
plane. Specifically, the amplitude of the field along $y=0$ decays
from the near-field plate ($z=0$) to the focal plane ($z=L$) by an
amount equal to:
\begin{equation}
e^{-q_0 L}q_0 L \label{eq: eq10}
\end{equation}
The fields $f(y,z=0)$ and $f(y,z=L)$, given by Equations (\ref{eq:
eq9}) and (\ref{eq: eq5}) respectively, are plotted in Figure
\ref{Figure3} for the case where $q_0 =10k_\circ$ and
$L=\lambda/16$.
%\begin{figure}
%\begin{center}
%\includegraphics[height= 6cm]{Figure11c.eps}
%\end{center}
%\caption{Electric field profiles at the focal plane ($z=L$) and at
%the surface of the near-field focusing plate ($z=0$). Note that the
%electric fields at the focal plane have been multiplied by 5 for
%clarity.} \label{Fig11}
%\end{figure}
From Equation (\ref{eq: eq5}), it can also be found that the
null-to-null beamwidth of the image at the focal plane ($z=L$) is:
\begin{equation}
\Delta y= 2\pi/q_0
 \label{eq: eq11}
\end{equation}
Expressing $q_0$ as a multiple of the free-space wavenumber $q_0=R_e
k_\circ$, the null-to-null beamwidth of the image can be rewritten
as:
\begin{equation}
\Delta y = \frac{2\pi}{R_e k_\circ}= \frac{\lambda_\circ}{R_e}
\label{eq: eq12}
\end{equation}
where $\lambda_\circ$ is the wavelength in free space, and $R_e$ is
what has often been referred to as the resolution enhancement
\cite{SmithRes, GrbicRes}. Further, expressing the distance to the
focal plane $L$ as a fraction of a free-space wavelength $L=
\lambda_\circ /M$, the decay of the field (Equation (\ref{eq:
eq10})) along the $z$ axis from the near-field plate to the focal
plane can be rewritten as:
\begin{equation}
e^{-2 \pi R_e / M} \frac{2 \pi R_e}{M} \label{eq: eq13}
\end{equation}
From Equation (\ref{eq: eq13}) it can be concluded that the ratio of
$R_e/M$ cannot be excessively high for the signal to still be
detectable at the focal plane of the near-field plate.

\section{Designing a Near-Field Focusing Plate}

Equation (\ref{eq: eq9}) indicates that the field at the near-field
plate exhibits both phase and amplitude variation. A simple way to
generate such a field distribution is to illuminate a reactance
sheet located at $z=0$ from the $-z$ direction with a plane wave.
The sheet should have a surface reactance that is a function of
position $y$ corresponding to the phase and amplitude variation of
the field.

To see how such a reactance sheet (near-field plate) can be
designed, let us first consider the simplified problem of
transmission through a sheet with uniform surface impedance. For
normal incidence, the transmission coefficient through a uniform
sheet is:
\begin{equation}
T=\frac{2Z_{sheet}}{\eta + 2Z_{sheet}} \label{eq: eq14}
\end{equation}
where $\eta=120 \pi \;\Omega$ is the impedance of free-space, and
$Z_{sheet}$ is the surface impedance of the reactance sheet. If the
surface impedance of the sheet is low ($Z_{sheet} \ll \eta$), the
transmission coefficient through the sheet can be approximated as:
\begin{equation}
T \approx \frac{2Z_{sheet}}{\eta} \label{eq: eq15}
\end{equation}
For this special case, the transmitted field is low, but its
magnitude and phase ($0^\circ$ or $180^\circ$) can be accurately
controlled. For example, if the reactance of the sheet is inductive
(of the form $j\omega L$), the transmission coefficient through the
sheet has a phase of $+90^\circ$, while if it is capacitive (of the
form $1/(j\omega C)$) the phase is $-90^\circ$. As a result,
capacitive and inductive surface impedances can be used to produce
fields that are $180^\circ$ out of phase. One can also change the
magnitude of the transmitted field by varying the magnitude of the
inductive or capacitive sheet reactance. Therefore, by using a sheet
with a reactance that is modulated as a function of $y$, one can
synthesize various field profiles including the one given in
Equation (\ref{eq: eq9}).

From the above discussion, it is clear that the reactance of an
impedance sheet can be manipulated to control the transmitted
electromagnetic field. Now let us consider the design of a
near-field plate (a specific type of impedance sheet) that focuses
energy from a plane wave to subwavelength dimensions at the $z=L$
plane. The plane wave is assumed to be polarized along the $x$
direction and normally incident from the $-z$ direction onto the
near-field plate located at $z=0$. The $y-$dependent surface
impedance of the near-field plate will be represented as
$Z_{sheet}(y)$. Similarly, the $x$-directed current density induced
on the near-field plate will be represented as $J_{x}(y)$. The
boundary condition along the reactance sheet (near-field plate) can
then be represented as a Fredholm integral equation of the second
kind:
\begin{equation}
E_\circ - \frac{k_\circ \eta}{4}\int^{W/2}_{-W/2}
J_{x}(y')H_0^{(2)}(k_\circ |y-y'|)dy'= Z_{sheet}(y)J_{x}(y)
\label{eq: eq16}
\end{equation}
where $E_\circ$ is the amplitude of the incident plane wave at
$z=0$, $H_\circ^{(2)}$ is a Hankel function of the second kind of
order zero, and $W$ is the width of the near-field plate. In the
integral equation the unknown current density appears both inside
and outside of the integral sign. The total field at the surface of
the near-field plate therefore is:
\begin{equation}
E_{total}(y) = E_\circ - \frac{k_\circ \eta}{4}\int^{W/2}_{-W/2}
J_{x}(y')H_0^{(2)}(k_\circ |y-y'|)dy' \label{eq: eq17}
\end{equation}
Equating $E_{total}(y)$ to the field desired at the surface of the
near-field plate, given by Equation (\ref{eq: eq9}), one can solve
for $J_x(y)$. Equation (\ref{eq: eq9}) has been multiplied by the
scaling factor $jK_\circ E_\circ$ to obtain the following equation
for $J_x(y)$:
\begin{equation}
\frac{jK_\circ E_\circ L[L \cos(q_0 y)+y \sin(q_0 y)]}{[L^2+y^2]} =
E_\circ - \frac{k_\circ \eta}{4}\int^{W/2}_{-W/2}
J_{x}(y')H_0^{(2)}(k_\circ|y-y'|)dy' \label{eq: eq18}
\end{equation}
The desired field has been multiplied by the imaginary number $j$ in
order to obtain predominantly passive (inductive and capacitive)
surface impedances for the near-field plate design. The variable
$K_\circ$ represents the amplitude of $E_{total}$ as a multiple of
the incident field $E_\circ$. A larger $K_\circ$ represents a higher
field amplitude at the surface of the near-field plate ($z=0$), and
therefore a more highly resonant plate. To obtain the unknown
current density $J_x(y)$, Equation (\ref{eq: eq17}) can be solved
numerically using the method of moments.  Finally, dividing
$E_{total}(y)$ by the computed current distribution $J_x(y)$, the
surface impedance $Z_{sheet}(y)$ can be found. Once the surface
impedance is found the design of the near-field plate is complete.

The procedure for deriving $Z_{sheet}(y)$ does not ensure that
$Z_{sheet}(y)$ is passive. To enforce that the lens is entirely
passive, only the imaginary part of the derived $Z_{sheet}(y)$ is
taken. The current density $J_{x}(y)$ is then solved for again by
plugging the passive $Z_{sheet}(y)$ into Equation (\ref{eq: eq16}).
Once the current density is found for the passive near-field plate,
the fields scattered by the near-field plate are computed using the
two dimensional free-space Green's function:
\begin{equation}
E_{x}^{s}(y,z)=  -\frac{k_\circ \eta}{4}\int^{W/2}_{-W/2}
J_{x}(y')H_0^{(2)}(k_\circ|\sqrt{(y-y')^2 + z^2}|)dy' \label{eq:
eq19}
\end{equation}
The total field at any point is then the sum of the incident
plane-wave $E_\circ e^{-jk_\circ z}$ and the scattered field
$E_{x}^{s}(y,z)$ due to the induced current density $J_{x}(y)$ on
the near-field plates.

Now let us consider a specific near-field plate design at 1.0 GHz
($\lambda_\circ =300 \; \mbox{mm}$). For this particular design $q_0
=10k_\circ =10 \omega /c$, or equivalently $\Delta y = \lambda_\circ
/10=30 \; \mbox{mm}$. In addition, the focal plane is chosen to be
$L= \lambda_\circ /16=18.7 \; \mbox{mm}$ from the near-field plate.
Hence, the near-field plate is capable of creating a focal spot with
a null-to-null beamwidth of $\lambda_\circ /10$ at a distance
$\lambda_\circ /16$ from the plate. In addition, the width of the
near-field plate ($W$) is chosen to be approximately $2$ wavelengths
in the $y$ direction and the constant $ K_\circ$ is set to $K_\circ
=6$. In other words, the field at $(y,z)=(0,0)$ is six times the
amplitude of the incident plane wave ($E_\circ$).

The current density $J_x(y)$ on the near-field plate is discretized
into 79 segments in order to solve Equation (\ref{eq: eq18})
numerically. The segments are centered at positions $(y,z)=(n\delta,
0)$, where $n$ is an integer from -39 to 39, and $\delta$ is the
width of each segment. The variable $\delta$ is chosen to be
$<\Delta y$ to mimic a continuous variation in surface impedance:
$\delta = \lambda_\circ /40=7.5\mbox{mm}$. Collocation (the point
matching method) \cite{Harrington} was used to solve for the current
density on the near-field plate, from which the surface impedance of
the near-field plate was subsequently found. In the computations it
was assumed that the incident plane wave is equal to $E_\circ
=1\mbox{V/m}$ at the surface of the near-field plate. Table
\ref{table1} shows the surface impedances of the $\delta$ segments
comprising the near-field plate. Since the plate is symmetric, the
surface impedances of only 40 segments ($n=0$ to $n=39$) are shown.
Column two of Table 1 shows the impedances that are derived directly
from Equation (\ref{eq: eq18}), while those in column 3 are the
passive surface impedances used in the design of the passive
near-field plate. They are completely imaginary and thus represent
inductive and capacitive surface impedances.

\begin{figure}
\begin{center}
\includegraphics[width= 8cm]{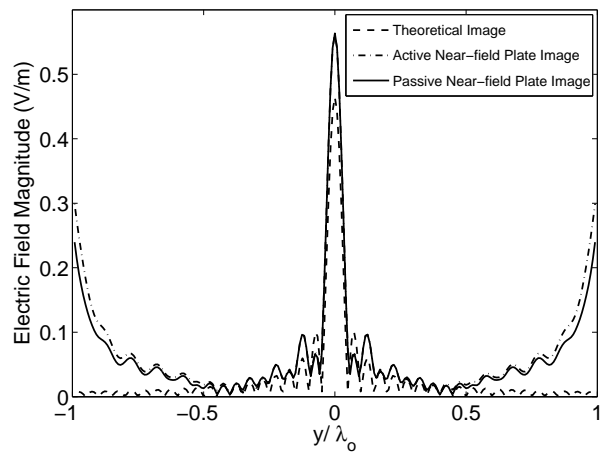}
\end{center}
\caption{Images formed by the near-field focusing plate: theory vs.
fullwave simulation results.} \label{Figure4}
\end{figure}

\begin{figure}
\begin{center}
\includegraphics[width= 8cm]{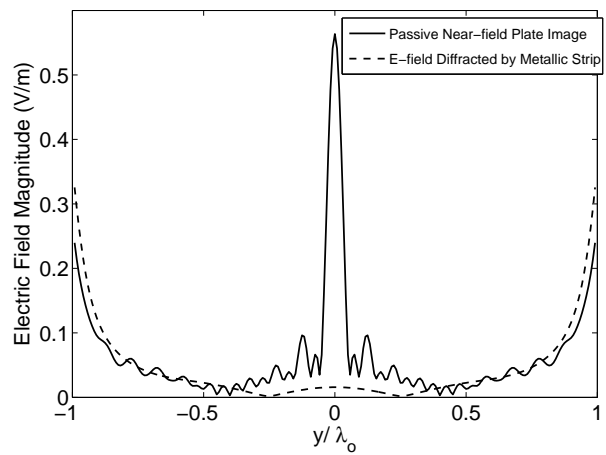}
\end{center}
\caption{Diffraction from a metallic strip and a near-field focusing
plate that are two wavelengths wide.} \label{Figure5}
\end{figure}

Plotted in Figure \ref{Figure4} are different electric field
profiles at the focal plane. The dotted line shows the theoretically
predicted image, which is simply a plot of Equation (\ref{eq: eq5})
multiplied by the constant $K_\circ=6$. The dash-dot line represents
the image that would be produced by the near-field plate possessing
the surface impedances given in column 2 of Table \ref{table1}. This
active near-field lens possesses reactances as well as positive
(loss) and negative (gain) resistive elements. Finally, the solid
line represents the image formed by the passive near-field focusing
plate. The active and passive plate images have a mainlobe that is
$\Delta y = \lambda/10$. The difference between the two images is
minimal, and they are both quite close to the theoretically
predicted image (dotted line). The images of the active and passive
plates, however, possess an increase in field magnitude near $y=\pm
1 \lambda$. This rise in field magnitude is due to the diffraction
of the incident plane wave from the edges of the near-field focusing
plate. Figure \ref{Figure5} compares the electric field diffracted
by a metallic strip that is two wavelengths wide to the electric
field diffracted by the near-field plate of the same width. As can
be seen from the plot, the electric field diffracted by the metallic
strip follows the field diffracted by the near-field focusing plate
near $y=\pm 1 \lambda$. This plot supports the fact that the rise in
electric field magnitude in Figure \ref{Figure5} is due to
diffraction. On the other hand, the electric field around $y=0$ is
quite different since the near-field focusing plate manipulates the
evanescent spectrum to create a sharp image, while the metallic
strip does not.
\section{Loss Performance}
\begin{figure}
\begin{center}
\includegraphics[width= 8cm]{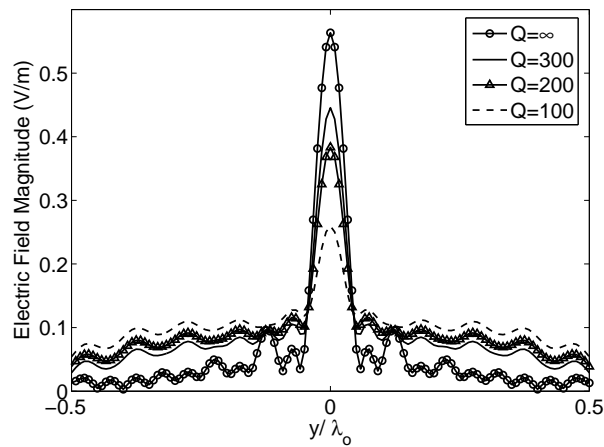}
\end{center}
\caption{The performance of a lossy near-field focusing plate.
Images are shown for near-field plates with varying quality factors,
Q.} \label{Figure6}
\end{figure}

Near-field focusing is a resonant phenomenon which degrades with
increased losses. In order to study the performance of a practical
near-field focusing plate, loss was added to the purely reactive
surface impedances of the passive plate given in column 3 of Table
1. The loss associated with a reactance is typically expressed in
terms of the quality factor, $Q$, which is defined as the ratio of
the surface reactance $X_{sheet}$(imaginary surface impedance) to
the surface resistance $R_{sheet}$ (real surface impedance):
\begin{equation}
Q =
\frac{X_{sheet}}{R_{sheet}}=\frac{\mbox{imag}(Z_{sheet})}{\mbox{real}(Z_{sheet})}
\label{eq: eq20}
\end{equation}
Figure \ref{Figure6} shows the focus for near-field plates with
various quality factors. For each graph, all surface impedances were
assigned the same quality factor. The plots show that the central
peak of the focus decreases and the sidelobes increase with
increasing loss. However, it is encouraging that the degradation of
the focus is gradual. For a printed metallic near-field focusing
plate at frequencies of a few gigahertz, quality factors of a couple
hundred can be expected. For such values of Q, the near-field
focusing is still very prominent. In practice, the inductive surface
impedances could be implemented as inductively loaded metallic
strips/wires while the capacitive surface impedances could be
implemented as capacitively loaded strips or metallic patches
printed on a microwave substrate. At optical frequencies, the
inductive surface impedances could be implemented using
nanofabricated plasmonic structures and the capacitive surface
impedances using dielectric structures \cite{Nader}.
\section{Conclusion}
The intrinsic properties and design of near-field focusing plates
have been described. These plates are planar structures that have
the ability to focus electromagnetic waves to subwavelength
dimensions. Moreover, a procedure has been outlined for designing a
near-field plate to achieve a desired image. Full-wave simulations
at microwave frequencies, have clearly demonstrated the near-field
plate's ability to overcome the diffraction limit. The effect of
losses on the performance of near-field focusing plates has also
been addressed. At microwave frequencies, near-field focusing plates
may find use in non-contact sensing and microwave imaging
applications. At optical frequencies, applications of this
technology may include lithography, microscopy and near-field
optical data storage.

\begin{table}
\vspace{1pt} \scriptsize \centering \caption{Surface Impedances of
the Near-Field Focusing Plate at 1 GHz} \vspace{1pt}
\begin{tabular}{ c c c} \hline
$n$ & $Z_{sheet}$ & Passive $Z_{sheet}$ \\
\hline 0   &    -0.0540 -24.8811i  &            -24.8811i  \\
1   &     -0.3078 -34.9762i &            -34.9762i  \\
2   &      0.0497 -18.8400i &            -18.8400i  \\
3   &      0.0830 -22.1429i &            -22.1429i  \\
4   &     -0.0866 -17.4066i &            -17.4066i  \\
5   &     -0.0874 -21.3768i &            -21.3768i  \\
6   &      0.0968 -14.1181i &            -14.1181i  \\
7   &      0.0705 -18.3300i &            -18.3300i  \\
8   &     -0.2215 -18.0665i &            -18.0665i  \\
9   &     -0.0862 -20.3643i &            -20.3643i  \\
10  &      0.1165 -12.1702i &            -12.1702i  \\
11  &      0.0536 -17.3865i &            -17.3865i  \\
12  &     -0.3545 -22.2030i &            -22.2030i  \\
13  &     -0.0480 -20.5220i &            -20.5220i  \\
14  &      0.0419 -10.3653i &            -10.3653i  \\
15  &      0.0049 -16.8255i &            -16.8255i  \\
16  &      0.3176 -33.9907i &            -33.9907i  \\
17  &      0.0482 -20.9474i &            -20.9474i  \\
18  &     -0.0994 - 8.6930i &            - 8.6930i  \\
19  &     -0.0696 -16.3809i &            -16.3809i  \\
20  &     27.6488 -97.2412i &            -97.2412i  \\
21  &      0.1922 -21.4451i &            -21.4451i  \\
22  &     -0.2328 - 7.2934i &            - 7.2934i  \\
23  &     -0.1455 -16.0315i &            -16.0315i  \\
24  &     29.1460 +63.0469i &            +63.0469i  \\
25  &      0.3231 -21.8837i &            -21.8837i  \\
26  &     -0.2867 - 6.2742i &            - 6.2742i  \\
27  &     -0.1752 -15.7837i &            -15.7837i  \\
28  &      5.6876 +26.9105i &            +26.9105i  \\
29  &      0.2934 -22.2145i &            -22.2145i  \\
30  &     -0.1685 - 5.5034i &            - 5.5034i  \\
31  &     -0.0641 -15.6296i &            -15.6296i  \\
32  &     -0.1209 +17.9582i &            +17.9582i  \\
33  &     -0.2211 -22.4108i &            -22.4108i  \\
34  &      0.3371 - 4.9484i &            - 4.9484i  \\
35  &      0.4372 -15.5174i &            -15.5174i  \\
36  &     -5.3426 + 9.7995i &            + 9.7995i  \\
37  &     -2.4484 -22.4780i &            -22.4780i  \\
38  &      1.7231 - 3.3621i &            - 3.3621i  \\
39  &      4.4763 -13.2700i &            -13.2700i  \\
\hline
\end{tabular}
\label{table1}
\end{table}%

\newpage

\bibliographystyle{IEEEtran}
\bibliography{references}
\end{document}